%% file: ms.tex
\documentclass[reprint, superscriptaddress, nofootinbib, aps, prc, floatfix]{revtex4-1}
\input{def.tex}

\newif\ifcolorfigs
\colorfigstrue        

\makeatletter
\def\ps@pprintTitle{%
 \let\@oddhead\@empty
 \let\@evenhead\@empty
 \def\@oddfoot{}%
 \let\@evenfoot\@oddfoot}
\makeatother

\begin{document}

\input{frontpage.tex}
\input{introduction}
\input{production_crosssection}
\input{experiment}
\input{crosssection}
\input{cosmic_activation}
\input{alternate_production}
\input{discussion}
\input{acknowledgements}
\bibliography{references}

\end{document}

%% file: def.tex
\usepackage[colorlinks=true,pdfstartview=FitV,linkcolor=blue,citecolor=blue,urlcolor=blue,bookmarks=false]{hyperref}
\usepackage[usenames,dvipsnames]{color}
\usepackage[sc]{mathpazo}
\usepackage[below]{placeins}
\usepackage{afterpage}
\usepackage[separate-uncertainty,retain-explicit-plus, per-mode = symbol, detect-weight=true, range-units=brackets]{siunitx}
\usepackage{longtable}
\usepackage{multirow}
\usepackage{rotating}
\usepackage[displaymath, mathlines]{lineno}
\usepackage{graphicx}
\usepackage{paralist}
\usepackage{wrapfig}
\usepackage{appendix}
\usepackage{vruler}
\usepackage{etoolbox}
\usepackage{tikz}
\usepackage{listings}
\usepackage{amsmath}
\usepackage{subfigure}
\usepackage{relsize}
\usepackage{etoolbox}
\usepackage{makecell}

\modulolinenumbers[5]
\newcommand{\arts}{$^{37}$Ar}
\newcommand{\artn}{$^{39}$Ar}
\newcommand{\arft}{$^{40}$Ar}
\newcommand{\isotope}[2]{$^{#2}${#1}}
\newcommand{\ntwon}{\mbox{(n, 2n)}}
\newcommand{\ntwong}{\mbox{(n, 2n$\gamma$)}}

\newcommand{\ncomblong}{[\mbox{(n, np)} + \mbox{(n, pn)} + \mbox{(n, d)}]}
\newcommand{\ncomb}{\mbox{(n, d)}}
\newcommand{\nfourn}{\mbox{(n, 4n)}}
\newcommand{\ngamma}{\mbox{(n, $\gamma$)}}

\newcommand{\uar}{UAr}
\newcommand{\aar}{AAr}
\newcommand{\ulbpc}{ULBPC}

\DeclareSIUnit\kgar{kg_\text{Ar}}
\DeclareSIUnit\bqkg{Bq/\kgar}
\DeclareSIUnit\atom{atom}
\DeclareSIUnit\atoms{atoms}
\DeclareSIUnit\atomsartn{atoms (^{39}Ar)}
\DeclareSIUnit\atomsarts{atoms (^{37}Ar)}
\DeclareSIUnit\atomscltn{atoms (^{39}Cl)}
\DeclareSIUnit\neutrons{neutrons}
\DeclareSIUnit\muons{\micro^-}
\DeclareSIUnit\gpg{g/g}
\DeclareSIUnit\c{$c$}
\DeclareSIUnit\second{sec}
\DeclareSIUnit\day{day}
\DeclareSIUnit\week{w}
\DeclareSIUnit\year{yr}
\DeclareSIUnit\standard{std}
\DeclareSIUnit\str{sr}



%% file: frontpage.tex
\title{Cosmogenic production of \artn~and \arts~in argon}


\newcommand{\pnnl}{Pacific Northwest National Laboratory, Richland, Washington 99352, USA}
\newcommand{\lanl}{Los Alamos National Laboratory, Los Alamos, New Mexico 87545, USA}

\author{R. Saldanha}\email{richard.saldanha@pnnl.gov}\affiliation{\pnnl}
\author{H. O. Back}\affiliation{\pnnl}
\author{R. H. M. Tsang}\affiliation{\pnnl}
\author{T. Alexander}\affiliation{\pnnl}
\author{S. R. Elliott}\affiliation{\lanl}
\author{\\S. Ferrara}\affiliation{\pnnl}
\author{E. Mace}\affiliation{\pnnl}
\author{C. Overman}\affiliation{\pnnl}
\author{M. Zalavadia}\affiliation{\pnnl}

\begin{abstract}
We have experimentally determined the production rate of \artn~and \arts~from cosmic ray neutron interactions in argon at sea level. Understanding these production rates is important for argon-based dark matter experiments that plan to utilize argon extracted from deep underground because it is imperative to know what the ingrowth of \artn~will be during the production, transport, and storage of the underground argon. These measurements also allow for the prediction of \artn~and \arts~concentrations in the atmosphere which can be used to determine the presence of other sources of these isotopes.  Through controlled irradiation with a neutron beam that mimics the cosmic ray neutron spectrum, followed by direct counting of \artn~and \arts~decays with sensitive ultra-low background proportional counters, we determined that the production rate from cosmic ray neutrons at sea-level is expected to be \SI{759 \pm 128}{\atoms\per\kgar\per\day} for \artn, and \SI{51.0 \pm 7.4}{\atoms\per\kgar\per\day} for \arts. We also performed a survey of the alternate production mechanisms based on the state-of-knowledge of the associated cross-sections to obtain a total sea-level cosmic ray production rate of \SI{1048 \pm 133}{\atoms\per\kgar\per\day} for \artn, \SI{56.7 \pm 7.5}{\atoms\per\kgar\per\day} for \arts~in underground argon, and \SI{92 \pm 13}{\atoms\per\kgar\per\day} for \arts~in atmospheric argon. \\
\end{abstract}

\keywords{argon, cosmogenic, activation, \artn, \arts}

\maketitle

%% file: introduction.tex
\section{Introduction}
Argon is a widely used medium for the detection of ionizing radiation. It has a high scintillation and ionization yield, allows for the propagation of scintillation photons and ionization electrons over large distances, and can be easily purified to remove non-noble impurities. Argon is therefore employed as the active medium for large neutrino detectors~\cite{rubbia2011underground, abi2018dune}, scintillation vetos~\cite{agostini2017background}, and direct-detection dark matter experiments \cite{benetti2008first, agnes2018darkside, amaudruz2018first}. Argon is particularly attractive for dark matter detectors as the time profile of the scintillation light enables pulse-shape discrimination (PSD) of signal-like nuclear recoils from radiogenic electron recoils.

Argon is the third-most abundant gas in the Earth's atmosphere, comprising roughly 0.93\% of the atmosphere by volume. Argon extracted from the atmosphere consists primarily of the stable isotopes \arft, \isotope{Ar}{36},  and \isotope{Ar}{38}. However, due to interactions of cosmic rays, atmospheric argon also contains three long-lived radioactive isotopes: \artn, \arts, and \isotope{Ar}{42}. The abundances and specific activity of the different isotopes in the atmosphere are given in Table~\ref{tab:ar_isotopes}. In this paper we will focus on the two radioisotopes that, due to their high specific activity, are most relevant for argon-based dark matter experiments: \arts~and \artn. 

\begin{table}[tbp]
\centering
\setcellgapes{1pt} \makegapedcells \renewcommand\theadfont{\normalsize\bfseries}%
\begin{tabular}{lrr} 
\hline
Isotope & Abundance & Specific Activity \\
&  & [\si{\bqkg}]\\   
\hline
\arft & 0.9960& Stable \\
\isotope{Ar}{36} & 0.0033 & Stable \\
\isotope{Ar}{38} & 0.0006 & Stable \\
\artn &  \num{8.2e-16} & \num{1.0}~\cite{loosli1968detection, benetti2007measurement}\\
\arts & $\approx$ \num{1.3E-20} & $\approx$ \num{4.5e-2} \cite{purtschert2017ar37}\\
\isotope{Ar}{42} & \num{6.8e-21} & \num{6.8e-5}~\cite{barabash2016concentration, agostini2014background}\\
\hline
\end{tabular}
\caption{Stable and long-lived isotopes of argon, along with their typical abundances~\cite{bohlke2014variation} and specific activity in atmospheric argon.}
\label{tab:ar_isotopes}
\end{table}

\artn~is a pure $\beta$-emitter with an end-point of 565 keV and a half-life of 268 years \cite{chen2018nuclear}. For large or low background argon-based detectors \artn~is often the dominant source of interactions at low energies. The $\beta$ decays of \artn~limit the sensitivity to rare events and can also create difficulties through signal pileup and high data acquisition rates. In dark matter detectors PSD is extremely effective at reducing the \artn~background at high energies, but at low energies the discrimination power is limited by the detected photon statistics and the energy threshold is often determined by the \artn~rate~\cite{agnes2015first, amaudruz2018first}.

To mitigate the effects of \artn, the next generation of argon-based dark matter detectors propose to use argon extracted from deep underground. The DarkSide-50 collaboration has demonstrated that underground argon (\uar) they are using as their dark matter target  has an \artn~rate of \SI{7.3e-4}{\bqkg}~\cite{agnes2016results}, a factor $\approx1400$ below atmospheric levels. The use of \uar~rather than atmospheric argon (\aar) has allowed for a reduction in energy threshold and increase in nuclear recoil acceptance while maintaining a background-free WIMP dark matter search~\cite{agnes2016results, agnes2018darkside}. Additionally, the use of low radioactivity \uar~is critical for low-mass dark matter searches which extend to lower energy thresholds than the standard WIMP search~\cite{agnes2018low}. The need for argon with low-levels of \artn~grows as future dark matter experiments move towards tonne-scale target masses and beyond. The relative background contributions of radioactive contaminants in external components will decrease due to self-shielding of the argon and a decreasing surface-to-volume ratio, making \artn~(which scales with the target mass) the likely dominant background. 

Similar to the commercial production of argon from the atmosphere, UAr must be extracted from crude naturally occurring gases underground. In the case of the DarkSide-50 experiment this crude gas is primarily CO$_2$, extracted from a gas well in Cortez, Colorado, USA \cite{back2012first}. The processing, transport, and storage of the UAr on the surface of the Earth exposes the UAr to cosmic rays until it can be suitably shielded underground.  Given the extremely stringent requirements on the levels of \artn~for dark matter detectors, cosmogenic production of \artn~in the \uar~is an important concern. 

\arts~decays purely through electron capture, producing low energy x-rays and Auger electrons, with a relatively short half-life of 35.01 days~\cite{cameron2012nuclear}. Since low background experiments are typically operated deep underground, shielded from cosmic rays, the \arts~activity typically decays below measurable levels within a few months, though the x-ray peak can be used as a low energy calibration source during early data-taking \cite{agnes2018low}. \arts~can also be produced by underground nuclear explosions through neutron reactions with calcium in the soil \cite{aalseth2011measurement}. Elevated rates of \arts~in the environment are a strong indicator of a nuclear explosion and can therefore be used to verify compliance with nuclear test ban treaties. Thus the natural cosmogenic production of \arts~in the atmosphere and surface soil gas acts as a background to underground nuclear monitoring \cite{riedmann2011natural}.

\begin{figure}[tbp]
   \centering
   \includegraphics[width=\linewidth]{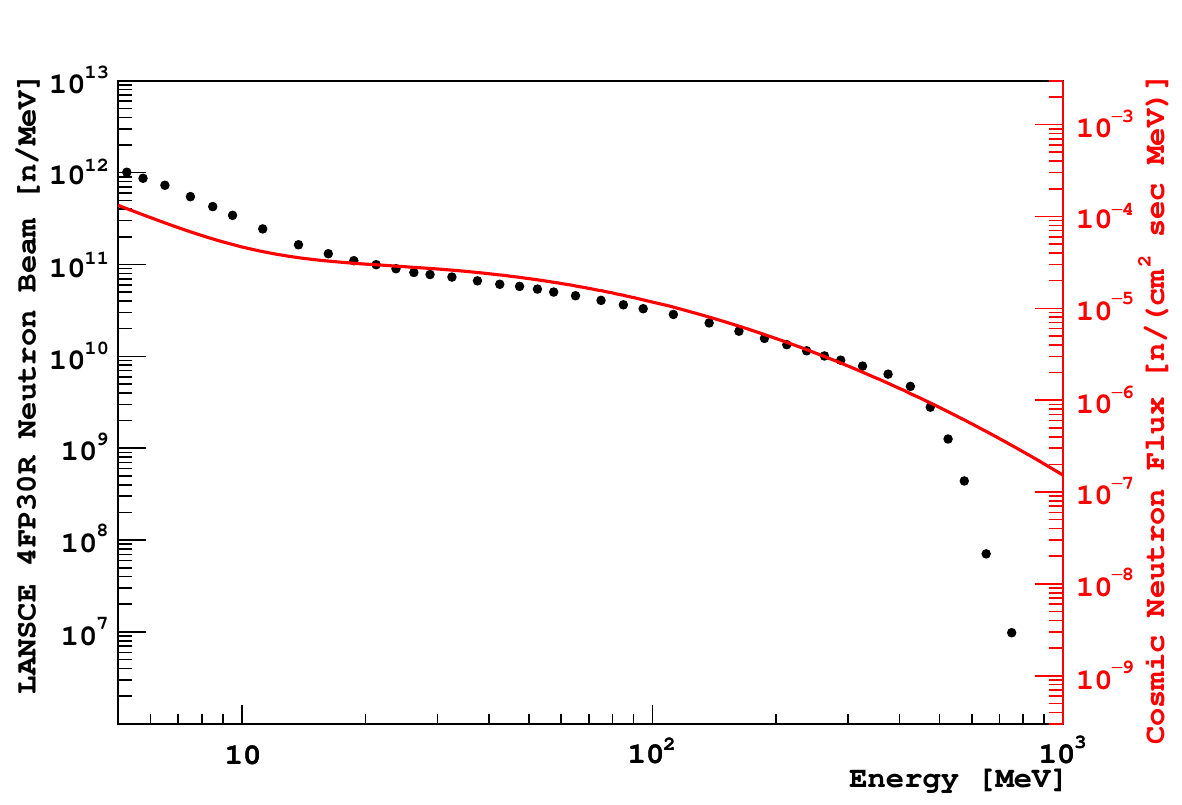} 
   \caption{Comparison of the LANSCE 4FP30R neutron beam with sea-level cosmic ray neutrons. The black data points and left vertical axis show the number of neutrons measured by the fission chamber during the 3-day beam exposure used for this measurement. The red continuous line and the right vertical axis show the reference cosmic ray neutron flux at sea-level for New York City during the mid-point of solar modulation \cite{gordon2004measurement}. }
   \label{fig:neutron_flux}
\end{figure}
	
The cosmogenic production of both \arts~and \artn~in argon is expected to be dominated by neutron-induced reactions but there are no existing measurements of the total production cross-section or integrated cosmogenic production rate. Estimates for dark matter experiments have therefore had to rely on semi-empirical calculations, which as we show, significantly underestimate the production rates. In this paper we describe a measurement of the integrated production rate from a neutron beam at the Los Alamos Neutron Science Center (LANSCE) ICE HOUSE facility \cite{lisowski2006alamos, icehouse} which has a very similar energy spectrum to that of cosmic ray neutrons at sea-level (see Figure~\ref{fig:neutron_flux}). While the spectral shape is similar, the neutron beam has a flux that is $\approx 4.5\times10^8$ times higher above 10 MeV, which allows for the production of measurable amounts of \artn~in short periods of time. We irradiated samples of underground and atmospheric argon for 3 days on the neutron beam and then measured the resulting activity in ultra-low background gas proportional counters (\ulbpc) at Pacific Northwest National Laboratory (PNNL). The ULBPCs are custom-built to detect low levels of \artn~and \arts~in gas samples for the purpose of radiometric dating and support of future nuclear test ban treaty monitoring \cite{aalseth2009design}, with a sensitivity to \artn~at the level of \SI{2.5e-2}{\bqkg}~\cite{hall201639, mace2018sensitivity} and to \arts~at the level of \SI{4.3e-3}{\bqkg}~\cite{aalseth2011measurement, mace2018sensitivity}. The high intensity of the LANSCE neutron beam and the sensitivity of the \ulbpc~detectors allow us to experimentally measure the \artn~and \arts~production rate and extrapolate them to estimate the sea-level cosmogenic neutron production rate. 

%% file: production_crosssection.tex
\section{$^{39}$A\lowercase{r} Production Mechanisms}

\begin{figure*}[t]
   \centering
   \begin{minipage}{0.49\textwidth}
   \centering
    \vspace{-5mm}
   \includegraphics[width=0.98\textwidth]{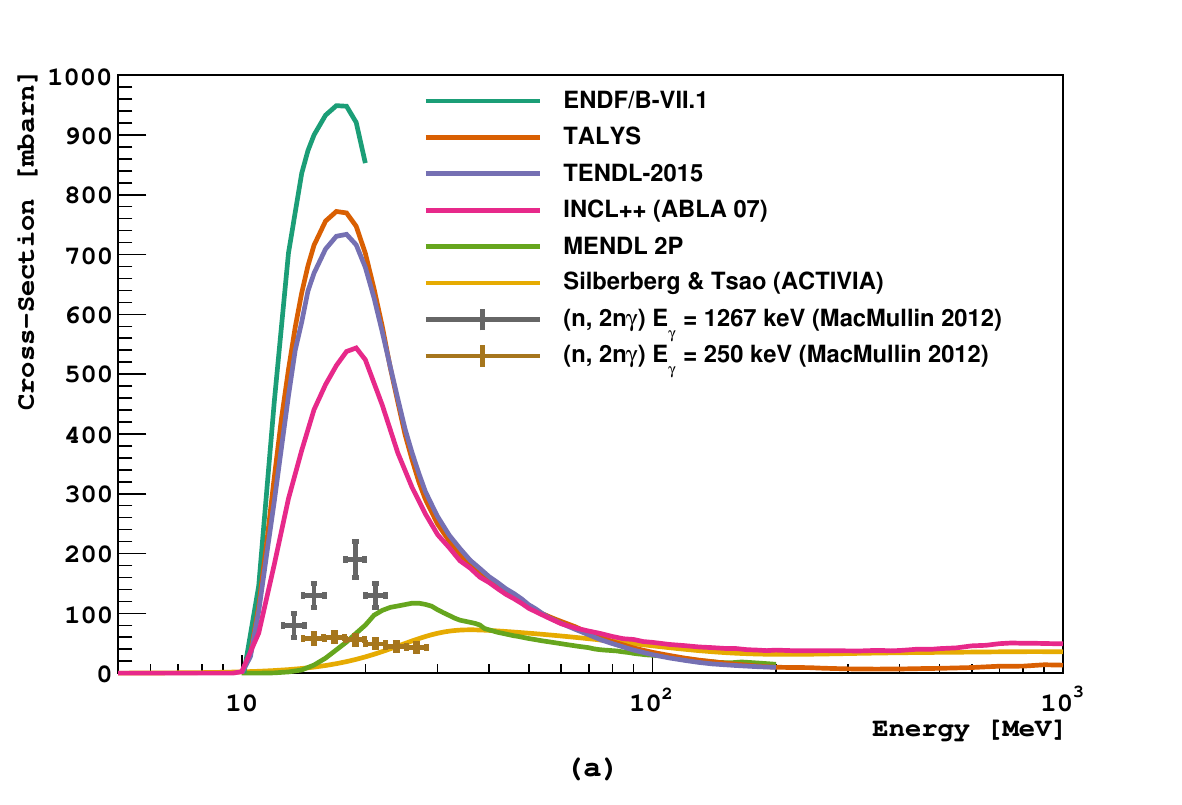} 
   \end{minipage}
   \begin{minipage}{0.49\textwidth}
   \centering
    \vspace{-5mm}
   \includegraphics[width=0.98\textwidth]{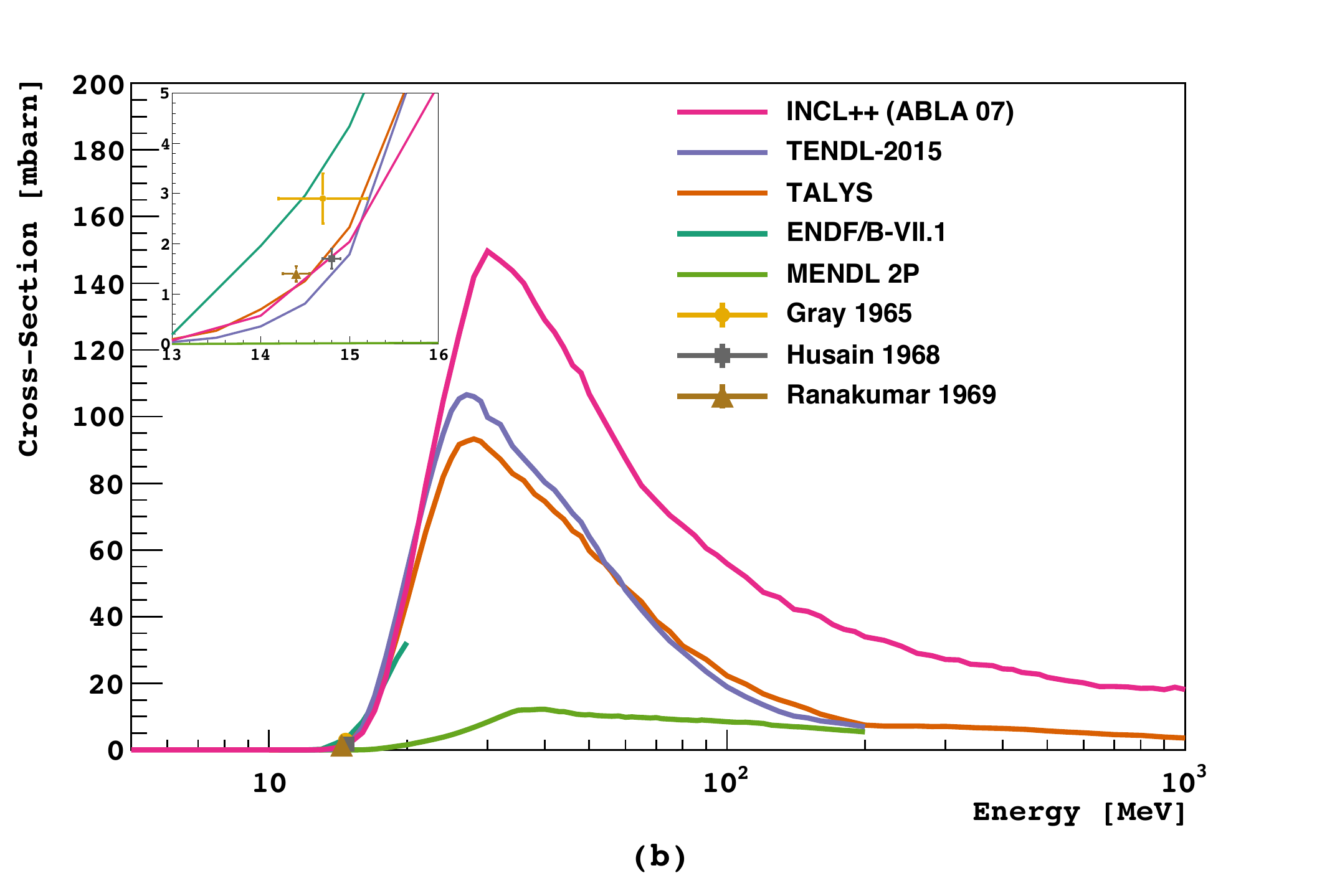} 
   \end{minipage}
   \caption{Estimates (continuous lines) and experimental measurements (data points) of \artn~production cross-sections from \arft. \hspace{\textwidth}
\textbf{(a)} \arft\ntwon\artn~cross-sections. Note that the experimental measurements are partial cross-sections to excited states of \artn~\cite{macmullin2012partial}.
\textbf{(b)} \arft\ncomb\isotope{Cl}{39} cross-sections with experimental measurements at $\approx 14.6$ MeV \cite{gray1965activation, husain196814, ranakumar1969thermal}. The Silberberg and Tsao \ncomb~cross-section calculated by ACTIVIA is combined with the \ntwon~cross-section shown in panel (a).}
   \label{fig:artn_cs}
\end{figure*}

The production of \artn~in the atmosphere is primarily due to cosmic rays. While several of the production channels do not have measured cross-sections, based on the known flux of the various cosmogenic particles and the general behavior of cross-sections for similar isotopes, it is estimated that interactions with fast neutrons account for more than 94\% of the total \artn~production in the atmosphere ~\cite{loosli1968detection}. There are two\footnote{The neutron-induced production of \isotope{S}{39}, which sequentially decays to \isotope{Cl}{39} and \artn, is estimated to be 4 orders of magnitude smaller than the direct \artn~production. For other possible production mechanisms see Section~\ref{sec:alt_prod}.} primary fast neutron reactions that result in the production of \artn: \arft\ntwon\artn~and \arft\ncomblong\isotope{Cl}{39}. In the latter case, which we will henceforth abbreviate to \arft\ncomb\isotope{Cl}{39}, the short-lived \isotope{Cl}{39} decays to \artn~through $\beta$ decay with a 55.6 minute half-life. Since \artn~is a pure $\beta$ emitter it is not possible to directly measure the total production cross-section by conventional methods that rely on $\gamma$-ray detectors to tag the reaction products. The only existing measurements are of the partial cross-sections \arft\ntwong\artn~to excited states of \artn~\cite{macmullin2012partial} and of \arft\ncomb\isotope{Cl}{39}~through the detection of gamma rays from the \isotope{Cl}{39}~decay~\cite{rama1961cosmic, gray1965activation, husain196814, ranakumar1969thermal}. 

Estimates of the cosmogenic activation rates have therefore had to rely on either semi-empirical calculations such as the Silberberg and Tsao equations \cite{silberberg1973partial, silberberg1973partial2, silberberg1977cross, silberberg1985improved, silberberg1990spallation, silberberg1998updated} that are employed by codes such as YIELDX \cite{tsao1996yieldx}, COSMO \cite{martoff1992cosmo}, and ACTIVIA \cite{back2008activia}, Monte Carlo simulations of the hadronic interactions between nucleons and nuclei that are performed by codes such as INCL \cite{boudard2013new}, ABLA \cite{kelic2008deexcitation}, TALYS \cite{koning2008talys}, etc., or compiled databases that combine calculations with experimental data such as ENDF \cite{chadwick2011endf}, MENDL-2P \cite{mclaughlin1998mendl}, and TENDL \cite{koning2012modern}. The \artn~production cross-section estimates from some of these different methods\footnote{The Silberberg and Tsao, and MENDL-2P cross-sections were obtained from the ACTIVIA code package \cite{activia2017}, the INCLXX cross-sections were calculated using the INCL++ code (v6.0.1) with the ABLA07 de-excitation model \cite{mancusi2014extension}, and the TALYS cross-sections calculated using TALYS-1.9 \cite{talys1.9}. The default parameters were used for all programs.}, along with the experimentally measured partial cross-sections, are shown in Figure~\ref{fig:artn_cs}. It can be seen that the estimates vary by up to an order of magnitude at certain energies, yielding similar sized variation in the predicted \artn~production rate. It should be noted that previous estimates of the cosmogenic activation rate of underground argon \cite{aalseth2018darkside} have used the Silberberg and Tsao semi-empirical calculations as implemented in the COSMO and ACTIVIA codes, which have the lowest predicted cross-sections and are significantly below the experimentally measured partial cross-sections. 

\begin{figure}[t]
   \centering
   \includegraphics[width=\linewidth]{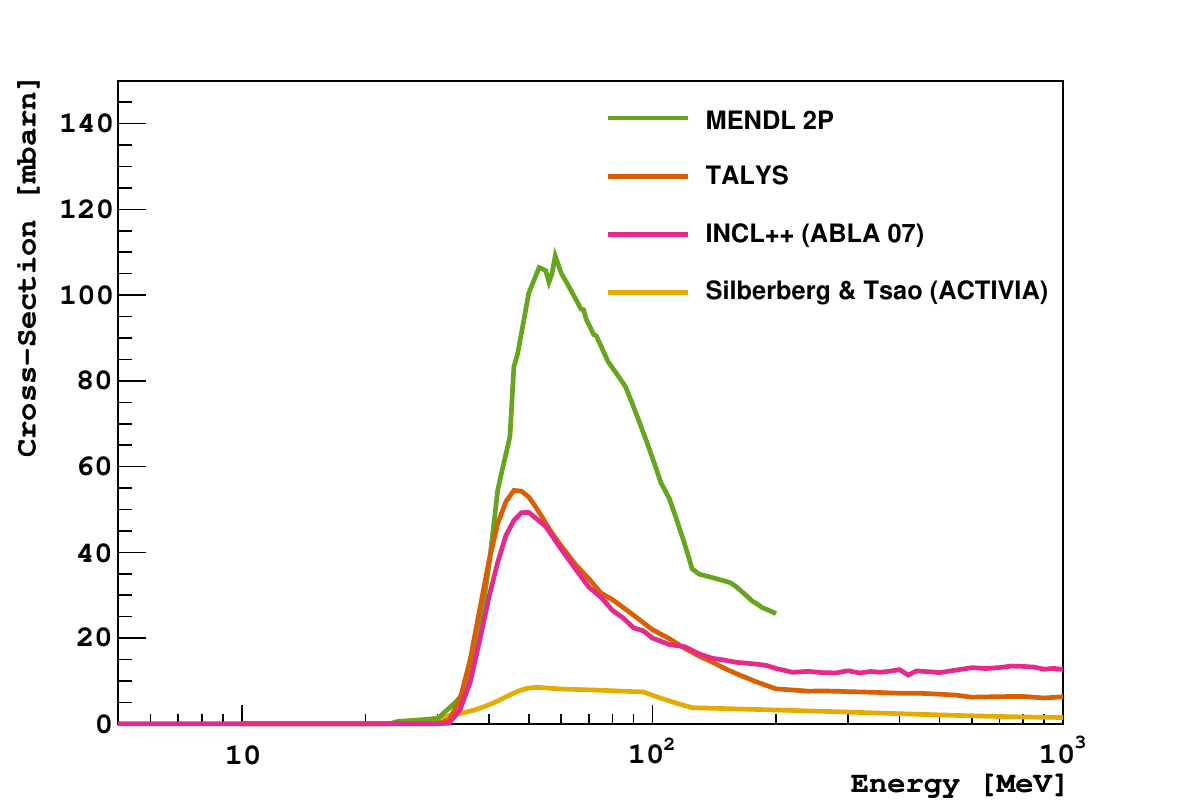} 
   \caption{Estimates of \arts~production cross-sections from \arft. There are no known experimental measurements.}
   \label{fig:arts_cs}
\end{figure}

\section{$^{37}$A\lowercase{r} Production Mechanisms}
The production of \arts~in the atmosphere is also dominated by neutron interactions on argon. Previous calculations \cite{loosli1970argon} estimate that 93\% of the total production in the troposphere is due to fast neutrons through \arft\nfourn\arts, with the remaining fraction\footnote{For other possible production mechanisms see Section~\ref{sec:alt_prod}.} due to the capture of thermal and epithermal neutrons through \isotope{Ar}{36}\ngamma\arts. The only detectable signal from \arts~decays are the low-energy x-rays and Auger electrons following the electron capture, and thus the \arft\nfourn\arts~production cross-section cannot be measured with $\gamma$-ray detectors. To our knowledge there are no known experimental measurements of this cross-section. Examples of estimates from semi-empirical calculations, simulations, and evaluated databases are shown in Figure~\ref{fig:arts_cs}. As with the \artn~production cross-sections, these estimates can vary by up to an order of magnitude.

%% file: experiment.tex
\section{Experiment}

\begin{figure}[t]
   \centering
   \includegraphics[width=0.9\linewidth]{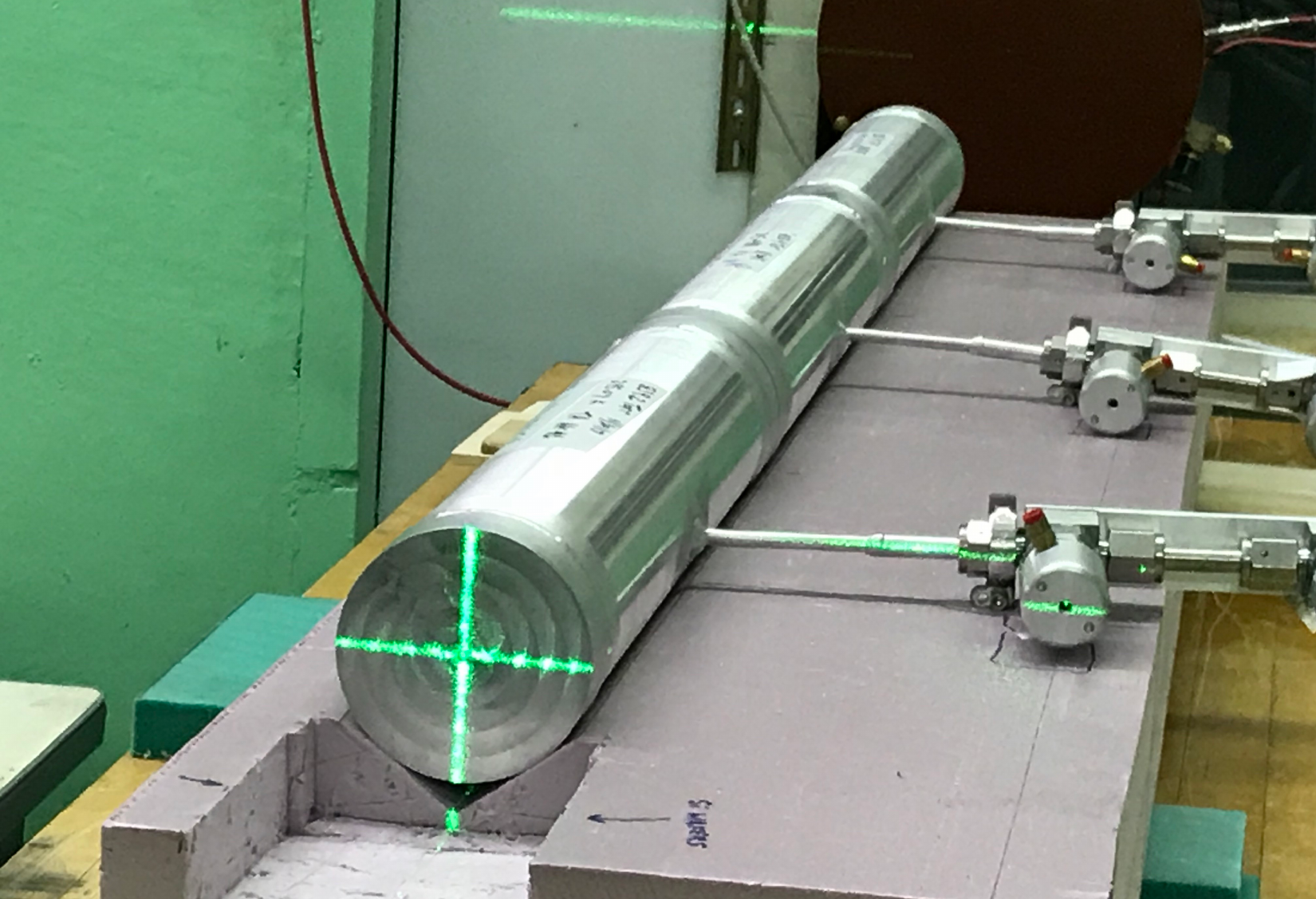} 
      \caption{Picture of the three argon gas cylinders placed on the Target 4 Flight Path 30 Right (4FP30R) at the LANSCE WNR ICE-HOUSE II facility. The beam direction is out of the page and the laser used for alignment of the cylinders can be seen on the face of the \aar~cylinder III, which is furthest downstream.}
   \label{fig:cyl_setup}
\end{figure}

Three samples of argon gas were prepared for irradiation, two \uar~samples and one \aar~sample. The \uar~samples were taken from the same source of gas that was used in the DarkSide-50 experiment and the \aar~sample was commercial ultra high purity (99.999\%)~grade argon obtained from Oxarc. An \aar~sample was used in addition to the \uar~samples to investigate if the different isotopic composition of the gases would lead to significantly different \artn~and \arts~production. In addition to \arft, \aar~contains \isotope{Ar}{36} at 0.334\% and \isotope{Ar}{38} at 0.063\% \cite{bohlke2014variation, meija2016isotopic}. \uar~is composed almost solely of \arft~with the \isotope{Ar}{36} concentration measured by mass spectrometry to be less than 0.01\%~and the \isotope{Ar}{38} concentration expected to be reduced by a similar factor compared to \aar, consistent with measurements of gas extracted from deep underground wells \cite{loosli1968detection}. For the sake of clarity, all numbers and uncertainties related to the gas samples in the main body of this paper refer to one of the \uar~samples (cylinder II). The corresponding numbers for the \aar~sample (cylinder III) are included in Table~\ref{tab:uncertainties}, while the other \uar~sample (cylinder I) was kept as contingency and not measured.

\begin{figure*}[t!]
   \centering
   \begin{minipage}{0.49\textwidth}
   \centering
    \vspace{-3mm}
   \includegraphics[width=0.98\textwidth]{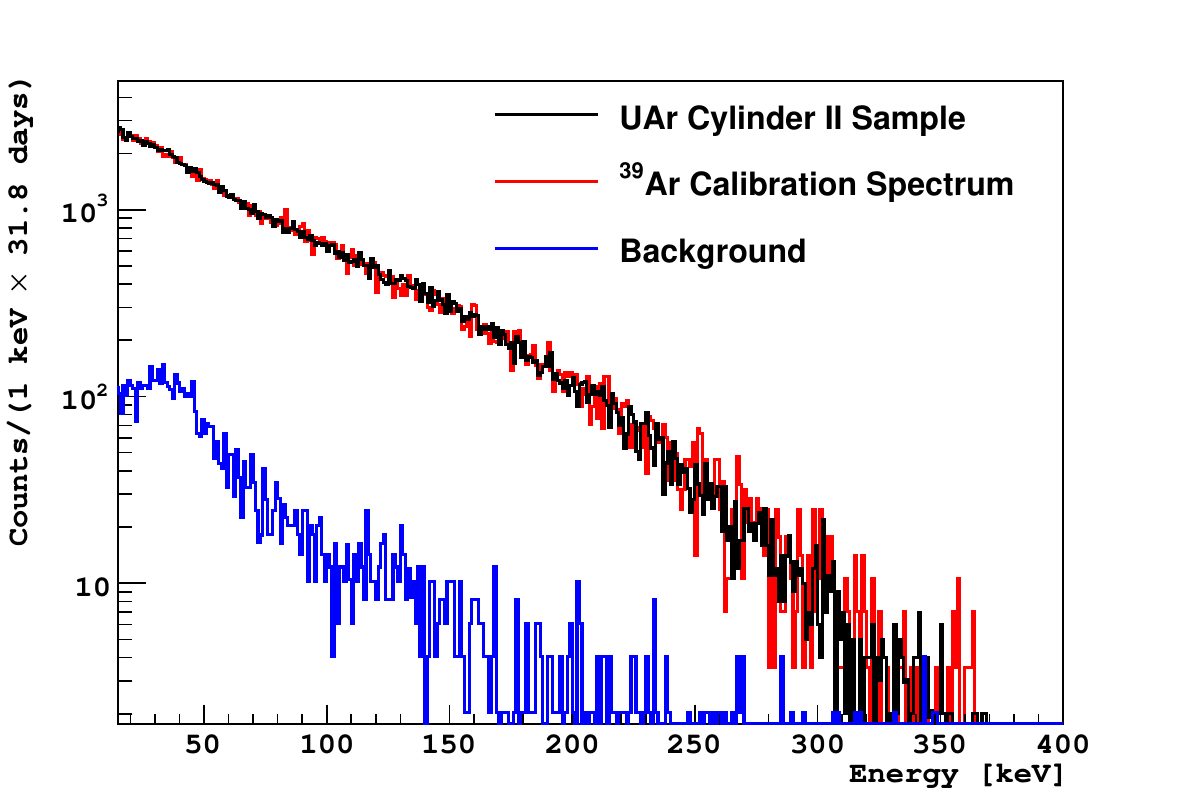} 
   \caption{Energy spectrum from gas in irradiated UAr cylinder II (black) as measured by the \ulbpc~at low gain. The spectrum from a \artn~calibration sample \cite{williams2017development} is overlaid for comparison (red) along with a background spectrum (blue).}
   \label{fig:ulbpc_artn}
   \end{minipage}
   \hfill
   \begin{minipage}{0.49\textwidth}
   \centering
   \includegraphics[width=0.98\textwidth]{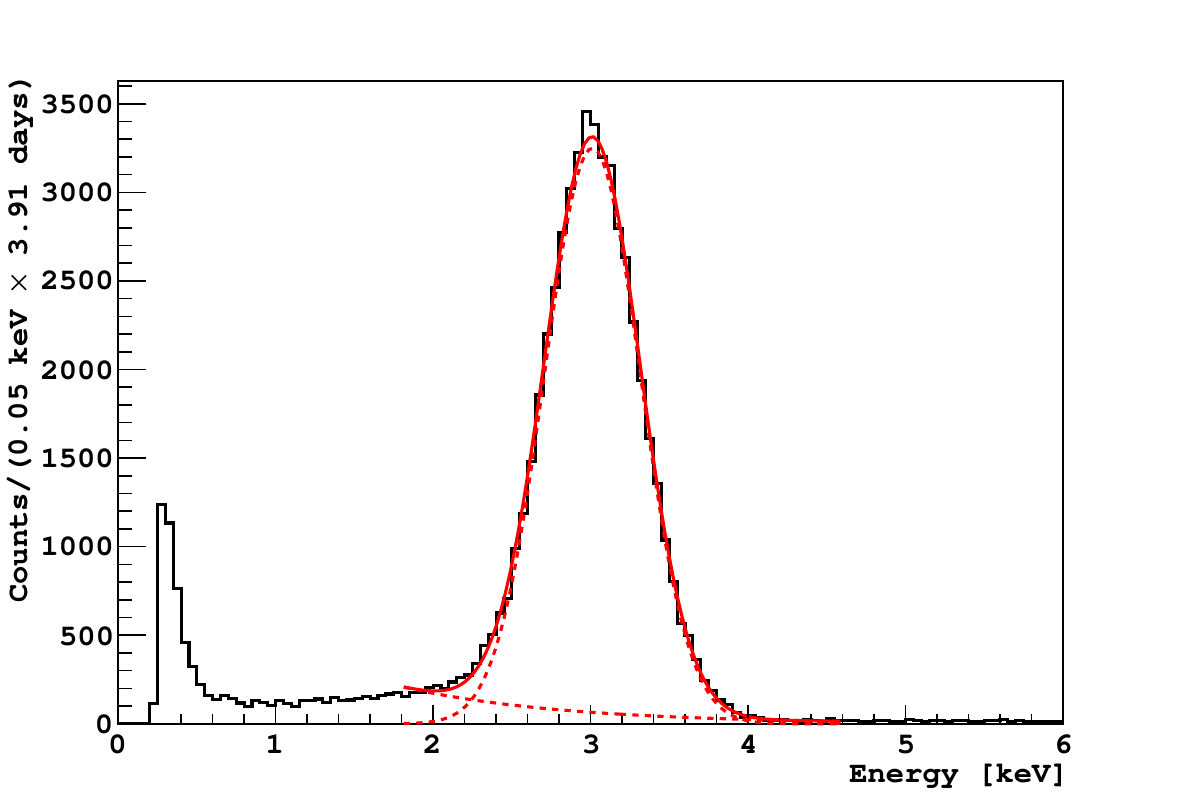} 
   \caption{Energy spectrum from gas in irradiated UAr cylinder II (black) as measured by the \ulbpc~at high gain. The peak region around 3 keV is fit (red solid) with a Gaussian (representing the \arts~Auger electrons and x-rays) and a falling exponential (for the underlying background).}
   \label{fig:ulbpc_arts}
   \end{minipage}
\end{figure*}

The gas samples were filled at \SI{1.12}{\bar} and \SI{296}{\kelvin} in three identical aluminum cylinders, each with an interior volume of \SI{1.96}{\liter} (calculated through calibrated volume-ratio measurements), corresponding to \SI{3.60 \pm 0.08}{\gram} of gas, where the uncertainty includes the uncertainties in pressure, temperature, and volume. The custom-designed cylinders had a nominal \SI{76.2}{\milli\meter} (3") internal diameter and \SI{439}{\milli\meter}~internal length and were fabricated from aluminum to minimize the attenuation of the neutron beam as well as activation of the cylinder material. The end caps of the cylinder, through which the neutron beam passed, had a thinned \SI{1.5 \pm 0.1}{\milli\meter}-thick central wall and were welded onto the cylindrical section. The gas was filled and emptied through a single \SI{6.35}{\milli\meter} diameter double-valved tube on each cylinder and the cylinders were evacuated to $\approx$ \SI{1e-6}{\bar} before being filled with the argon gas sample.

The samples were irradiated at the LANSCE WNR ICE-HOUSE II facility~\cite{icehouse} on Target 4 Flight Path 30 Right (4FP30R). A broad-spectrum (\numrange[range-phrase = --]{0.2}{800}~\si{\MeV}) neutron beam was produced via spallation of 800 MeV protons on a tungsten target. A 50.8 mm (2") diameter beam collimator was used to restrict the majority of the neutrons to within the diameter of the gas cylinder. The neutron fluence was measured with \isotope{U}{238} foils by an in-beam fission chamber \cite{wender1993fission} placed downstream of the collimator. The beam has a pulsed time structure which allows the incident neutron energies to be determined using a time-of-flight (TOF) measurement between the proton beam pulse and the fission chamber signals \cite{lisowski2006alamos, wender1993fission}. As shown in Figure~\ref{fig:cyl_setup}, the three gas cylinders were placed end-to-end with the two \uar~cylinders (I and II) closest to the fission chamber and \aar~cylinder III furthest away. The closest end of the \uar~cylinder I was located \SI{791}{\milli\meter} downstream from the fission chamber.  

The beam exposure took place over three days between 11$^{\text{th}}$-14$^{\text{th}}$~December 2017. The fluence measured by the fission chamber is shown in Figure~\ref{fig:neutron_flux}, with a total of \num{1.03 \pm 0.05 e13} neutrons above 10 MeV. The uncertainty is dominated by the systematic uncertainty in the \isotope{U}{238}(n, f) cross-section which varies from 2\% to 5\% below 200 MeV \cite{carlson1997238U} (which includes most of the production range) and we have conservatively assumed 5\% across the whole spectrum. The uncertainty in the neutron energy spectrum due to the timing uncertainty in the TOF measurement (1.2 ns) is negligible for this measurement. While the nominal beam diameter was set by the 2" collimator, the cross-sectional beam profile has significant tails at larger radii, with roughly 10\% falling outside the 3" diameter of the gas cylinders. Additionally the beam is slightly diverging, with an estimated cone opening angle of \SI{0.233}{\degree}. A Geant4~\cite{agostinelli2003geant4, allison2016recent} simulation that included the measured beam profile and beam divergence, the measured neutron spectrum, and the full geometry and materials of the gas cylinder, was used to calculate the neutron fluence through each cylinder. The estimated reduction factor of the neutron fluence through \uar~cylinder II in the 10-200 MeV range, compared to the fluence measured by the fission chamber, is \num{0.876 \pm 0.028}, where the uncertainty includes the uncertainty in the beam profile and a $\pm~0.25"$ uncertainty in the alignment of the beam center with the cylinder axes.

Following the irradiation the cylinders were stored for a cool-down period of roughly a month before they were shipped to PNNL for counting. The measured pressure of the argon in the irradiated cylinders upon return to PNNL matched the initial fill pressures within uncertainties, indicating that no argon had leaked out during the transportation or irradiation. Activation calculations indicated that along with \artn~and \arts, several other long-lived radioisotopes would also be produced by the irradiation, most notably \isotope{P}{32}, \isotope{P}{33}, \isotope{S}{35}, \isotope{Na}{22}, \isotope{H}{3}, and \isotope{Be}{7}. However previous experiments involving the irradiation of argon \cite{brodzinski1970high, reyss1981production} found that activation products remain in the irradiation canister when argon is transferred out. Since tritium is especially dangerous as it can introduce a long-lived radioactive background into the \ulbpc, the argon gas was cryopumped through a hydrogen getter into a smaller cylinder for loading into the \ulbpc. The irradiated gas was then mixed with 10\% methane to form P-10 count gas. 

A sample of the P-10 gas mixture from \uar~cylinder II was loaded into the \SI{100.5}{\cubic\centi\meter} \ulbpc~at \SI{6.44}{\bar} and \SI{295}{\kelvin}, corresponding to an argon mass of \SI{1.07}{\gram}. The irradiated \uar~gas was counted for \num{31.8} days at low gain to measure the \artn~spectrum from \numrange[range-phrase = --]{15}{400}~\si{\keV} and was also counted at high gain from \numrange[range-phrase = --]{0}{15}~\si{\keV} for \num{3.91} days to measure the \SI{3}{\keV} \arts~peak.

\begin{table*}[t!]
   \centering
      \begin{tabular}{c|cc|cc} 
      & \multicolumn{2}{c|}{\uar~cylinder II} & \multicolumn{2}{c}{\aar~cylinder III}\\
      & Value & Unc.[\%] & Value & Unc.[\%] \\ 
      \hline 
      \rule{0pt}{1.1\normalbaselineskip}
      Atomic Weight \cite{bohlke2014variation} &  $39.9624^{+0}_{- 0.003}$ & 7$\times10^{-3}$ & \num{39.94780 \pm 0.00002} & 3.8$\times10^{-5}$\\
      \arft~Isotopic Abundance \cite{bohlke2014variation} & $1.0000^{+0}_{- 0.0007}$ & 0.07 & \num{0.996035 \pm 0.000004} & 4.2$\times10^{-4}$\\
      Mass [\si{\gram}]  & \num{3.603 \pm 0.084} & 2.3 & \num{3.626 \pm 0.084} & 2.3\\
      \arft~Areal Density [\si{atoms\per\square\cm}] & \num{1.218 \pm 0.020 e21}& 1.6 &  \num{1.219 \pm 0.020 e21} & 1.6 \\
      Neutrons ($>$ 10 MeV) through cylinder  &  \num{9.03 \pm 0.54 e12} & 5.9 & \num{8.77 \pm 0.53 e12} & 6.0\\
       \hline  \rule[-1.1ex]{0pt}{0pt}
      \rule{0pt}{1.1\normalbaselineskip}
      \ulbpc~Sample Mass [\si{\gram}] & \num{1.070 \pm 0.011} & 1.0 & \num{1.052 \pm 0.011} & 1.0 \\
      \ulbpc~\artn~Activity [\si{\milli\becquerel}] & \num{46.12 \pm 0.46} & 1.0 & \num{43.96 \pm 0.46} & 1.0\\
      \ulbpc~\arts~Activity [\si{\milli\becquerel}] & \num{216.3 \pm 6.8} & 3.1 & \num{90.07 \pm 2.8} & 3.1\\
      \hline  \rule[-1.1ex]{0pt}{0pt}
      \rule{0pt}{1.1\normalbaselineskip}
      \artn~Half-life [yrs]  \cite{chen2018nuclear} & \num{268 \pm 8} & 3 & \num{268 \pm 8} & 3 \\
      Pre-existing \artn~Activity [\si{\milli\becquerel}]  & \num{4.3 \pm 0.4 e-3} & 9.5 & \num{3.7 \pm 0.3} & 8.5 \\
      Beam-Averaged Cross-Section [cm$^2$] & \num{1.72 \pm 0.12 e-25} & 7.2 & \num{1.67 \pm 0.12 e-25} & 7.3\\
      \textbf{Beam Induced \artn~Activity [\si{\milli\becquerel}]} & \textbf{\num{155.4 \pm 4.2}} &  \textbf{2.7} &  \textbf{\num{146.7 \pm 4.1}} &  \textbf{2.8}\\
      \hline
      \rule{0pt}{1.1\normalbaselineskip}
      \arts~Half-life [days]  \cite{cameron2012nuclear} & \num{35.011 \pm 0.019} & 0.054 & \num{35.011 \pm 0.019} & 0.054\\
      \arts~Decay Correction Factor  & \num{2.7996 \pm  0.0061 e-2} & 0.22 & \num{1.2205 \pm  0.0032 e-2} & 0.26\\
      Pre-existing \arts~Activity  [\si{\becquerel}] & \num{1.6 \pm 0.5 e-4} & 33 & \num{1.6 \pm 0.5 e-4} & 33 \\
      Beam-Averaged Cross-Section [cm$^2$] & \num{1.033 \pm 0.074 e-26}& 7.2 & \num{1.031 \pm 0.075 e-26} & 7.3 \\
       \textbf{Beam Induced \arts~Activity [\si{\becquerel}]} &  \textbf{\num{26.0 \pm 1.0}} &  \textbf{4.0} &  \textbf{\num{25.3 \pm 1.0}} &  \textbf{4.0} \\
   \end{tabular}
   \caption{Central values and uncertainties of the parameters and experimentally measured values used to determine the beam-induced \artn~and \arts~activity in \uar~cylinder II and \aar~cylinder III. See text for details. }
   \label{tab:uncertainties}
\end{table*}

The \artn~detection efficiency of the \ulbpc~detector was calibrated with a \SI{62.6}{\milli\becquerel} \artn~source \cite{williams2017development}, and a \SI{15.6}{\day} background run was also acquired. As shown in Figure~\ref{fig:ulbpc_artn}, the extremely good agreement in spectral shape between the efficiency spectrum and the data indicate that no other $\beta$-decaying activation products were present in the gas sample. After subtracting the measured background spectrum and accounting for the calibrated detection efficiency of \num{0.688 \pm 0.003}, the measured \artn~activity in the \ulbpc~was found to be \SI{46.1 \pm 0.5}{\milli\becquerel}. 

For \arts, the spectrum was fit with a Gaussian to represent the $\approx$ \SI{2.8}{\keV} Auger electrons and x-rays following a K-shell electron capture and an exponential decay for the underlying \artn~spectrum (shown in Figure~\ref{fig:ulbpc_arts}). The branching ratio for the K-shell electron capture is \SI{90.2 \pm 0.2}{\percent}~\cite{cleveland1998measurement, cameron2012nuclear} and the combined \ulbpc~fiducial volume efficiency for both Auger electrons and x-rays is $0.79 \pm 0.02$ \cite{williams2016development}. Including systematic uncertainties from the choice of fit range and response function we obtained a total measured \arts~activity in the \ulbpc~of \SI{216 \pm 7}{\milli\becquerel}, scaled to the start of the counting time (accounting for the decay of \arts~during the measurement period). 

Scaling these activities by the fraction of the total irradiated gas measured and the radioactive decay between the start of the irradiation and the measurement, we found an \artn~activity of \SI{155 \pm 4}{\milli\becquerel} and an \arts~activity of \SI{26.0 \pm 1.0}{\becquerel} for all the gas in the irradiated \uar~cylinder II. Following the same procedure described above, for the gas from \aar~cylinder III we found an \artn~activity of \SI{150 \pm 4}{\milli\becquerel} and an \arts~activity of \SI{25.3 \pm 1.0}{\becquerel}. 

One must also account for the cosmogenic \artn~and \arts~activity that is present in the samples even without the beam irradiation. Atmospheric argon was found to contain a specific activity of \SI{1.01 \pm 0.08}{\becquerel\per\kgar}~\cite{benetti2007measurement} of \artn, corresponding to an activity of \SI{3.7 \pm 0.3}{\milli\becquerel} in \aar~cylinder III, which must be subtracted from the measured activity when calculating the beam activation rate. The \uar~sample was obtained from the same source of gas used by DarkSide-50 that was found to have an \artn~specific activity of \SI{0.73 \pm 0.11}{\milli\becquerel\per\kgar}. Compared to the DarkSide-50 target which was transported deep underground in February 2015, our \uar~samples spent approximately 1170 additional days on the surface, exposed to cosmic-rays. Even assuming the highest considered cross-section, the total pre-existing rate of \artn~in the irradiated gas sample is roughly \SI{4.3}{\micro\becquerel}, negligible compared to the measured activity. 

For the short-lived \arts, both samples will have reached equilibrium activity from cosmogenic production at sea-level. Measurements of the \arts~activity in the low troposphere (excluding variations due to in-flows of stratospheric air, outgassing of soil air, and emissions from nuclear installations) indicate an equilibrium value of \SIrange[range-phrase = --, range-units = single]{0.5}{1.0}{\milli\becquerel\per\cubic\metre~air} (\SIrange[range-phrase = --, range-units = single]{30}{60}{\milli\bqkg}) \cite{purtschert2017ar37}. The equilibrium rate at sea-level is expected to be significantly lower than the rate averaged over the troposphere (in Section~\ref{sec:alt_prod} we estimate it to be (\SIrange[range-phrase = --, range-units = single]{0.7}{1.1}{\milli\bqkg})), but we conservatively use the averaged tropospheric value as an upper limit, corresponding to total activity of \SI{1.6 \pm 0.5 e-4}{\becquerel}, which is in any case negligible compared to the measured activity.

The final estimates for the beam-induced activity in the \uar~[\aar] sample at the end of the irradiation are \SI{155 \pm 4}{\milli\becquerel} [\SI{147 \pm 4}{\milli\becquerel}] \artn~and  \SI{26 \pm 1}{\becquerel} [\SI{25 \pm 1}{\becquerel}] \arts. The specific input values used as well as the included statistical and systematic uncertainties are listed in Table~\ref{tab:uncertainties}. To compare the results between the \uar~and \aar~samples we can divide the measured activity by the total number of neutrons and target \arft~atoms to obtain a beam-averaged cross-section (listed in Table~\ref{tab:uncertainties}). The ratio of the \uar~to \aar~beam-averaged cross-section, after eliminating common systematic uncertainties, is \num{1.029 \pm 0.055} and \num{1.002 \pm 0.052} for \artn~and \arts~respectively. The good agreement between the irradiated \uar~and \aar~activities indicate that, as expected, there was no appreciable contribution from neutron interactions with the small fractions of naturally occurring \isotope{Ar}{36} and \isotope{Ar}{38} isotopes present in the \aar~sample. The agreement also verifies the accuracy of the simulations of the beam attenuation between the targets and the independent \ulbpc~activity measurements. 

%% file: crosssection.tex
\section{Cross-Sections}
If the neutron beam had an energy spectrum identical to that of the cosmic ray neutron flux we could simply estimate the cosmogenic production rate by scaling the measured activity by the ratio of the cosmic-ray neutrons to that of the neutron beam. However the beam spectrum falls off faster at higher energies than that of cosmic rays (see Figure~\ref{fig:neutron_flux}). Thus we must rely on a model for the production cross-sections to extrapolate from the beam measurement to the cosmogenic production rate. 

\begin{table*}[t!]
   \centering
      \begin{tabular}{ccccc} 
      Cross-Section& Pred. LANSCE & Meas./Pred. LANSCE & Pred. Cosmogenic  & Scaled Cosmogenic \\
      Model & \artn~Activity & \artn~Activity & \artn~Prod. Rate & \artn~Prod. Rate\\
       & [mBq] & & [\si{\atoms\per\kgar\per\day}] & [\si{\atoms\per\kgar\per\day}]\\
         \hline
          Silberberg \& Tsao (ACTIVIA) & \num{37.1 \pm 2.5} & \num{4.19 \pm 0.31} & \num{200 \pm 25} & \num{840 \pm 120}\\
          MENDL-2P & \num{36.0 \pm 2.5} & \num{4.31 \pm 0.32} & \num{188 \pm 24} & \num{810 \pm 120}\\
          TENDL-2015 & \num{162 \pm 11} & \num{0.961 \pm 0.071} & \num{726 \pm 91} & \num{700 \pm 100}\\
          TALYS & \num{168 \pm 12} & \num{0.924 \pm 0.068} & \num{753 \pm 94} & \num{700 \pm 100}\\
          INCL++ (ABLA07) & \num{172 \pm 12} & \num{0.902 \pm 0.067} & \num{832 \pm 104} & \num{750 \pm 110}\\
         \hline
         \end{tabular}
   \caption{\arft\ntwon\artn~and \arft\ncomb\isotope{Cl}{39}~production rates for different cross-section models. The second and fourth column show the predicted rates for the LANSCE neutron beam exposure and sea-level cosmic ray neutrons respectively. The third column shows the ratio of the neutron beam prediction to the experimental measurement while the final column shows the cosmogenic production rate scaled by that ratio.}
   \label{tab:artn_production_rates}
\end{table*}

\begin{table*}[t!]
   \centering
      \begin{tabular}{ccccc} 
      Cross-Section& Pred. LANSCE & Meas./Pred. LANSCE & Pred. Cosmogenic  & Scaled Cosmogenic \\
      Model & \arts~Activity & \arts~Activity & \arts~Prod. Rate & \arts~Prod. Rate\\
       & [Bq] & & [\si{\atoms\per\kgar\per\day}] & [\si{\atoms\per\kgar\per\day}]\\
         \hline
          Silberberg \& Tsao (ACTIVIA) & \num{9.19 \pm 0.57} & \num{2.83 \pm 0.21} & \num{17.9 \pm 2.2} & \num{50.7 \pm 7.4}\\
          MENDL-2P & \num{79.7 \pm 4.9} & \num{0.326 \pm 0.024} & \num{155 \pm 19} & \num{50.5 \pm 7.3}\\
          TALYS & \num{39.1 \pm 2.4} & \num{0.666 \pm 0.049} & \num{76.8 \pm 9.6} & \num{51.1 \pm 7.4}\\
          INCL++ (ABLA07) & \num{39.9 \pm 2.5} & \num{0.653 \pm 0.048} & \num{79.3 \pm 9.9} & \num{51.8 \pm 7.5}\\
         \hline
         \end{tabular}
   \caption{\arft\nfourn\arts~production rates for different cross-section models (TENDL-2015 cross-sections were not available). The second and fourth column show the predicted rates for the LANSCE neutron beam exposure and sea-level cosmic ray neutrons respectively. The third column shows the ratio of the neutron beam prediction to the experimental measurement while the final column shows the cosmogenic production rate scaled by that ratio.}
   \label{tab:arts_production_rates}
\end{table*}

We can evaluate the accuracy of the different cross-section models by comparing the predicted \artn~and \arts~production rates from the LANSCE neutron beam irradiation to the measured rates. The number of isotopes atoms $N(t)$ at a given time $t$ is governed by the equation
\begin{linenomath*}
\begin{align}
\frac{dN}{dt} = + P(t) -\frac{N(t)}{\tau}
\end{align}
\end{linenomath*}
where $\tau$ is the mean life [\si{\second}] of the isotope decay and $P(t)$ is the isotope production rate [\si{\atoms\per \second}]. Ignoring any existing isotope concentration (subtracted off from the measured experimental value), the measured decay rate at any time after the start of the beam irradiation is given by
\begin{linenomath*}
\begin{align}
D(t) \equiv \frac{N(t)}{\tau} = \frac{e^{-\frac{t}{\tau}}}{\tau}\int\limits_0^t P(t')e^{\frac{t'}{\tau}}dt'
\end{align}
\end{linenomath*}
For a given cross-section model $\sigma(E)$ [cm$^2$]
\begin{linenomath*}
\begin{align}
P(t) = n_a \int S(E,t) \cdot \sigma(E)~dE
\end{align}
\end{linenomath*}
where $n_a$ is the areal number density of the target argon atoms [\si{\atoms\per \cm\squared}]~and $S(E,t)$ is the energy spectrum of neutrons [\si{\neutrons \per \MeV \per \second}]. The second column of Table~\ref{tab:artn_production_rates} shows the calculated values for the different \artn~cross-section models considered, with the corresponding numbers for \arts~shown in Table~\ref{tab:arts_production_rates}. We note that not all the cross-section models considered span the entire range of neutron energies. The TENDL 2015 and MENDL-2P cross-sections are only reported up to 200 MeV, while the TALYS cross-sections have been extended up to 1 GeV \cite{koning2014extension}. The INCL++ model can handle neutron-induced reactions up to 15-20 GeV, while the Silberberg \& Tsao semi-empirical cross-section calculations implemented in ACTIVIA can be performed at any energy, though the cross-sections are assumed to be independent of energy above $\approx 3$ GeV. 

From the ratio of the experimentally measured values to the predictions (shown in the third column of Table~\ref{tab:artn_production_rates} and ~\ref{tab:arts_production_rates}) it can be seen that the TENDL-2015, TALYS, and INCL++ models all predict \artn~activities within 10\% of the measured LANSCE activation value. However, the Silberberg and Tsao cross-section model used as the default by the ACTIVIA and COSMO codes, and previously used to estimate the cosmogenic production rate for argon dark matter experiments \cite{aalseth2018darkside}, predict values more than a factor of 4~smaller than the experimental measurement. The predictions for the \arts~production vary from the experimental measurements by roughly a factor three in both directions, with the TALYS and INCL++ models accurate to within about 50\%. 

If we assume that the shape of the cross-section model as a function of energy is correct, the ratio of the experimentally measured activity to the predicted activity represents the normalization factor that will need to be applied to each model to best match the experimental data. In the next section we will use this ratio to estimate the \arts~and \artn~production rates from cosmic ray neutrons. 

%% file: cosmic_activation.tex
\section{Production Rates from Fast Cosmic Ray Neutrons}

There have been several measurements and calculations of the cosmic ray neutron flux (for eg. \cite{hess1959cosmic, armstrong1973calculations, ziegler1996terrestrial}). The intensity of the neutron flux varies with altitude, location in the geomagnetic field, and solar magnetic activity (though the spectral shape does not vary as significantly) and correction factors must be applied to calculate the appropriate flux \cite{desilets2001scaling}. The most commonly used reference spectrum for sea-level cosmic ray neutrons is the so-called "Gordon" spectrum \cite{gordon2004measurement} (shown in Figure~\ref{fig:neutron_flux}), which is based on measurements at five different sites in the United States, scaled to sea-level at the location of New York City during the mid-point of solar modulation. We have used the parameterization given in \cite{gordon2004measurement} which agrees with the data to within a few percent. The spectrum uncertainties at high energies are dominated by uncertainties in the spectrometer detector response function ($<4$\% below 10 MeV and 10-15\% above 150 MeV) and we have assigned an average uncertainty of 12.5\%.

The production rate per unit target mass $P'$ [\si{\atoms\per\kgar\per\second}] of isotopes from the interaction of cosmic ray neutrons can be written as
\begin{linenomath*}
\begin{align}
P' = n \int \Phi(E) \cdot \sigma(E)~dE
\end{align}
\end{linenomath*}
where $n$ is the number of target atoms per kilogram of argon and $\Phi(E)$ is the cosmic neutron flux [\si{\neutrons\per\cm\squared\per\second\per\MeV}]. The integral is evaluated from 1 MeV to 10 GeV.

The predicted production rate per unit target mass for the cross-section models considered is shown in the fourth column of Table~\ref{tab:artn_production_rates} and Table~\ref{tab:arts_production_rates} for \artn~and \arts~ respectively. Scaling these values by the ratio of the experimental to predicted activities for the LANSCE neutron beam, we obtain our best estimates for the cosmic neutron induced production rates per unit target mass, shown in the final columns. The spread in the values for the different cross-sections is an indication of the systematic uncertainty in the extrapolation from the LANSCE beam measurement to the cosmic neutron spectrum. If either the LANSCE neutron beam spectral shape was the same as that of the cosmic ray neutrons, or the cross-section models all agreed in shape, the central values in the final column would be identical. The estimated activation rate from cosmic ray neutrons is \mbox{$(759 \pm 56_{exp} \pm 65_{cs} \pm 95_{nf})$ \si{\atomsartn\per\kgar\per\day}} and \mbox{$(51.0 \pm 3.8_{exp} \pm 0.6_{cs} \pm 6.4_{nf})$ \si{\atomsarts\per\kgar\per\day}}, where the first uncertainty listed is due to experimental measurement uncertainties (represented by the average uncertainty on the ratio of the measured to predicted activity from the LANSCE beam irradiation for a specific cross-section model), the second is due to the uncertainty in the shape of the cross-section models (calculated as the standard deviation of the scaled cosmogenic production rates by the different models), and the third is due to the uncertainty in the sea-level cosmic neutron flux. 

%% file: alternate_production.tex
\section{Alternate Production Mechanisms}
\label{sec:alt_prod}
\begin{figure*}[t!]
   \centering
   \begin{minipage}{0.49\linewidth}
   \centering
   \vspace{-3mm}
   \includegraphics[width=0.98\textwidth]{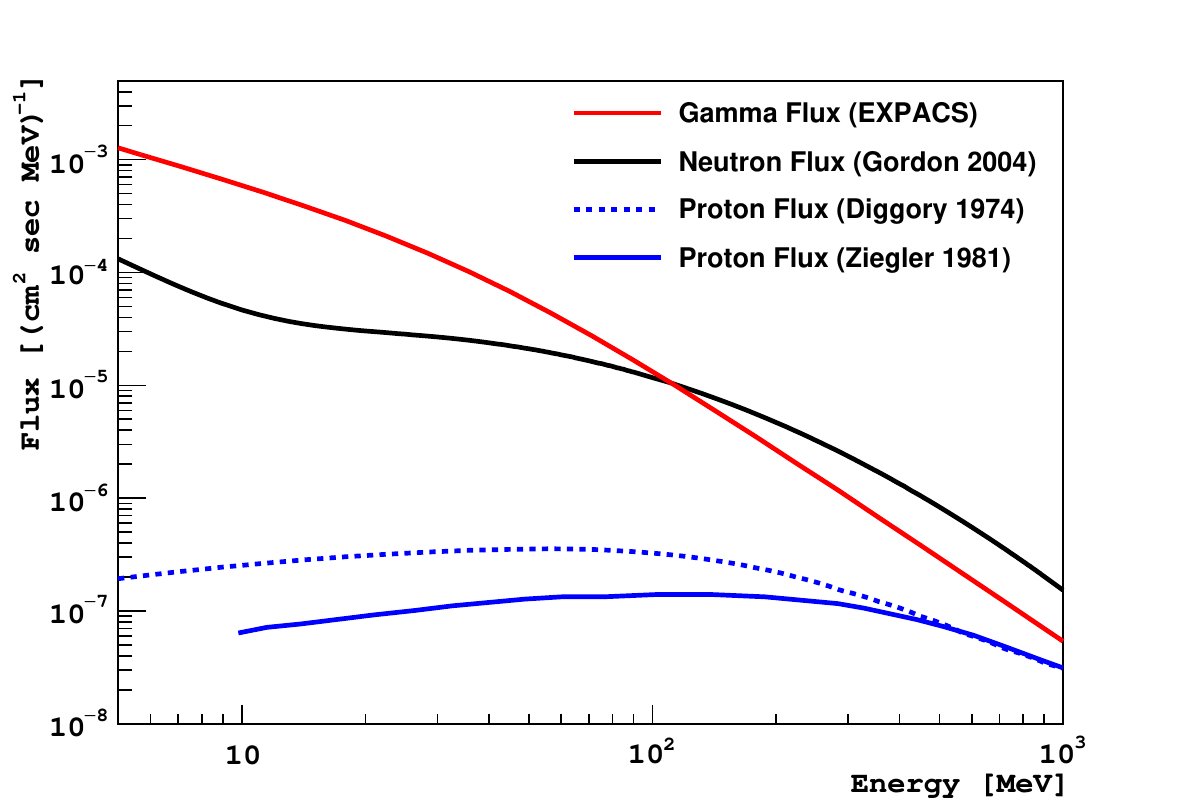}
   \caption{Sea-level cosmic ray flux of gamma-rays \cite{expacs}, neutrons \cite{gordon2004measurement}, and protons \cite{diggory1974momentum, ziegler1981background}.}
   \label{fig:alternate_production_flux}
   \end{minipage}
   \hfill
   \begin{minipage}{0.49\linewidth}
   \centering
   \includegraphics[width=0.98\textwidth]{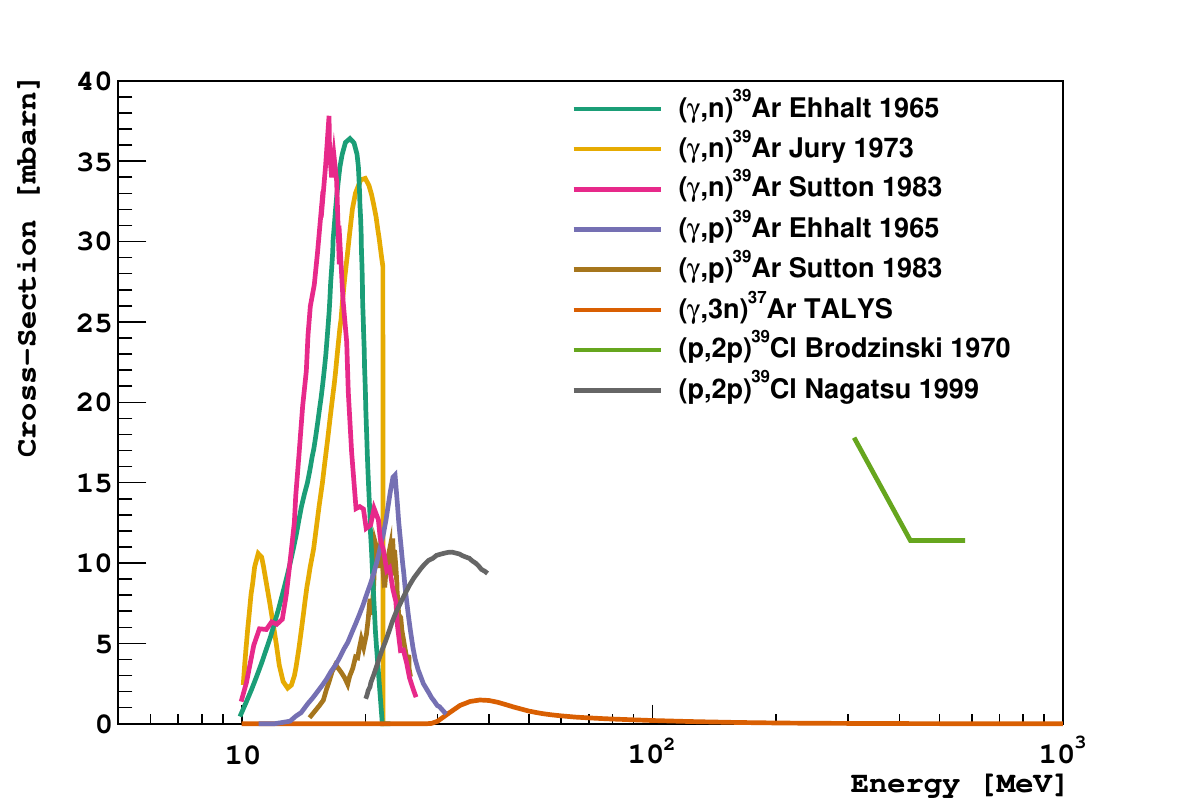}
    \caption{Production cross-sections for \artn~and \arts~on \arft~through gamma and proton interactions. See text for details and references.}
   \label{fig:alternate_production_cs} 
   \end{minipage}
\end{figure*}

\begin{figure}[t!]
   \centering
   \includegraphics[width=\linewidth]{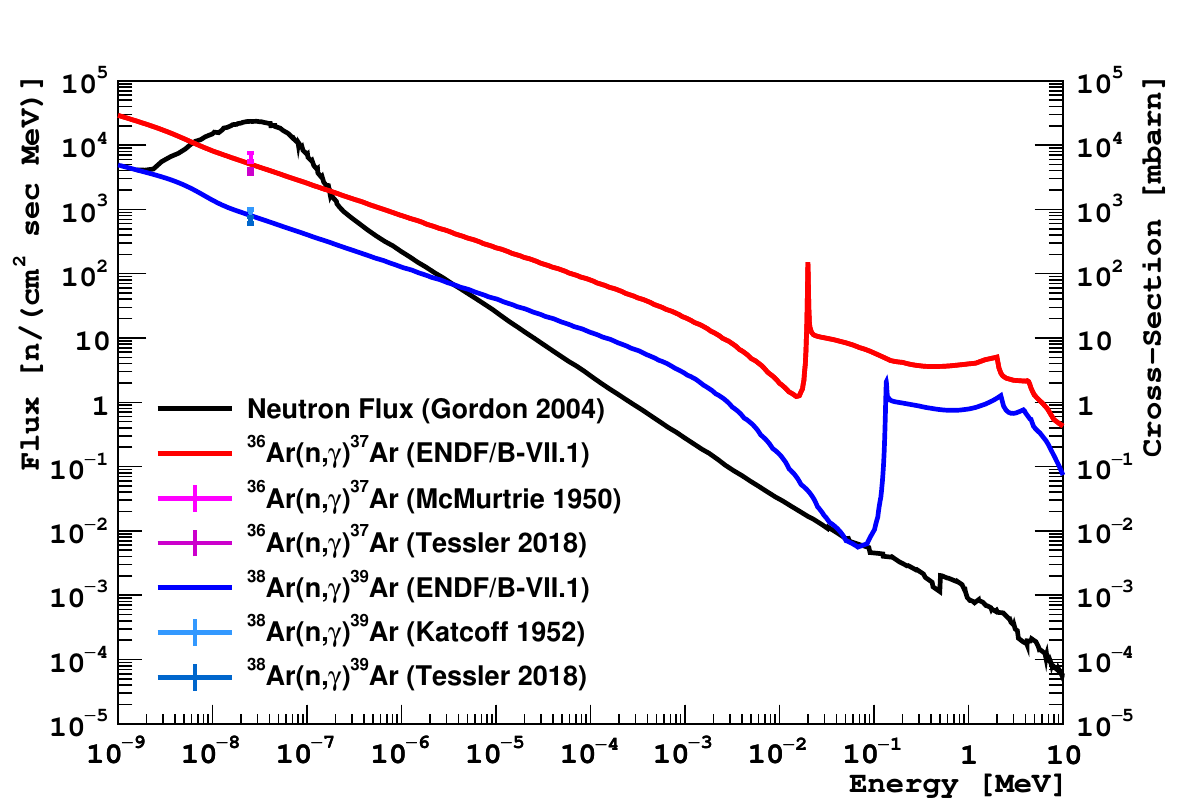} 
   \caption{Thermal and epi-thermal neutron flux (black, left axis) and capture cross-sections on \isotope{Ar}{36}~(red) and \isotope{Ar}{38}~(blue) along with experimental data points at thermal energies \cite{katcoff1952thermal, mcmurtrie1950thermal, tessler2018stellar}.}
   \label{fig:thermal_neutrons}
\end{figure}

In addition to activity induced by fast neutrons, interactions of gamma-rays, muons, protons, and thermal neutrons also contribute to the total production rate of \artn~and \arts. Previous calculations estimated that interactions of fast neutrons account for 94\% of the total \artn~production in the atmosphere \cite{loosli1968detection} and 93\% of the production of \arts~in the troposphere \cite{loosli1970argon}. However, at the Earth's surface the ratio of the fast neutron flux compared to other cosmogenic particles is significantly different due to different attenuation lengths in the atmosphere and the relative contributions must be re-evaluated. In the following two sub-sections we describe the methods we used to estimate the individual contributions using existing measurements and models.

\subsection {Alternate \artn~Production Mechanisms}
\subsubsection*{\arft($\mu$, n)\isotope{Cl}{39}}
The production rate from muon captures at sea-level, $P_{\mu0}$, can be estimated by
\begin{linenomath*}
\begin{align}
P_{\mu0} = R_0 \cdot \frac{\lambda_c(\text{\arft)}}{\lambda_d + \lambda_c(\text{\arft)}} \cdot f^*(\text{\isotope{Cl}{39}})
\end{align} 
\end{linenomath*}
where $R_0$ = \SI{484 \pm 71}{\muons\per\kgar\per\day} is the rate of stopped muons at sea-level \cite{charalambus1971nuclear} and we have added a 10\% uncertainty to account for the 10\% difference in the (Z/A) value for argon compared to air, $\lambda_c$(\arft) = \SI{1.20 \pm 0.08 e6}{\per\second} is the capture rate of muons on argon \cite{bertin1973nuclear}, $\lambda_d$ = \SI{4.552 e5}{\per\second} is the decay rate of muons \cite{tanabashi2018m}, and $f^*(\text{\isotope{Cl}{39}})$  = \num{0.490 \pm 0.014} is the effective probability of producing \isotope{Cl}{39} or \isotope{S}{39} \cite{klinskikh2008muon, parvu2018short}. The sea-level production rate from muon captures is therefore estimated to be \SI{172 \pm 26}{\atoms\per\kgar\per\day}.

\subsubsection*{\arft($\gamma$, n)\artn~and \arft($\gamma$, p)\isotope{Cl}{39}}
The flux of high energy $\gamma$~rays at the Earth's surface was obtained using the PARMA analytical model \cite{sato2016analytical} as implemented in the EXPACS software program \cite{expacs}. Similar to the neutron spectrum, we used New York city as our reference location for the $\gamma$ spectrum, which is shown in Figure~\ref{fig:alternate_production_flux}. We verified that the $\gamma$ flux predicted by the model agreed with experimental measurements \cite{ryan1979atmospheric} to within 20\% in the energy range of interest (10-30 MeV), which we used as our estimate of the systematic uncertainty. Experimental measurements of the \mbox{\arft($\gamma$, n)\artn}~cross-section \cite{ehhalt1965kernphotoeffekt, jury1973photoneutron, sutton1983photodisintegration} and the \mbox{\arft($\gamma$, p)\isotope{Cl}{39}}~cross-section \cite{ehhalt1965kernphotoeffekt, sutton1983photodisintegration} are shown in Figure~\ref{fig:alternate_production_cs}. Where multiple measurements exist we have used the mean and sample standard deviation of the calculated production rates as our estimates of the central value and uncertainty. The estimates for the individual processes are shown in Table~\ref{tab:artn_total_production_rates} with the total sea-level production rate from $\gamma$~rays estimated to be \SI{113 \pm 24}{\atoms\per\kgar\per\day}.

\subsubsection*{\arft(p, 2p)\isotope{Cl}{39} and \arft(p, pn)\artn}
At sea-level the flux of 10-100 MeV cosmic ray protons is at least thirty times lower than that of cosmic ray neutrons due to the additional electromagnetic interactions of protons in the atmosphere. To estimate the production rate from protons we have used the proton spectra from \cite{ziegler1981background} and \cite{diggory1974momentum} (scaled by the angular distribution from the EXPACS code) as shown in Figure~\ref{fig:alternate_production_flux}. Measurements of the \arft(p, 2p)\isotope{Cl}{39}~cross-section at low \cite{nagatsu1999excitation} and high \cite{brodzinski1970high} energies are shown in Figure~\ref{fig:alternate_production_cs} but, due to the low proton flux, the contribution to the overall production rate at sea-level is negligible. We are not aware of any measurements of the  \arft(p, np)\artn~cross-section (again probably due to the difficulty of detecting \artn) and have therefore based our estimates on the \arft\ncomb\artn~cross-section from TALYS, scaled by the same factor used in Table~\ref{tab:artn_production_rates}. As before, we have used the mean and sample standard deviation of the calculated production rates with the different proton spectra and cross-sections as our estimates of the central value and uncertainty, yielding a sea-level production rate of \SI{3.6 \pm 2.2}{\atoms\per\kgar\per\day}.

\subsubsection*{\isotope{Ar}{38}\ngamma\artn}
In atmospheric argon (AAr), \artn~can also be produced through the capture of thermal and epi-thermal neutrons on \isotope{Ar}{38}. For the low energy neutron flux we used the measurements taken in New York (see Figure~\ref{fig:thermal_neutrons} taken from Figure 4 in \cite{gordon2004measurement}). We note that even after correcting for altitude, location, and solar activity, the flux of low energy neutrons varies due to differences in the local environment \cite{gordon2004measurement} and we have therefore assumed a 30\% uncertainty in the overall normalization of the flux.  For the \isotope{Ar}{38}\ngamma\artn~cross-section we have used the ENDF/B-VIII.I values \cite{chadwick2011endf} (shown in Figure~\ref{fig:thermal_neutrons}) which agrees well with experimental data at thermal energies \cite{katcoff1952thermal, tessler2018stellar}. Due to the low abundance of \isotope{Ar}{38} the contribution to the total production rate (\SI{1.1 \pm 0.3}{\atoms\per\kgar\per\day}) is negligible.

The estimates for each of these alternate production mechanisms for \artn~is summarized in Table~\ref{tab:artn_total_production_rates}, with fast neutrons contributing 73\% of the total production rate of \artn~at sea-level. As partial verification, one can compare the estimates of the total \isotope{Cl}{39} production rate to an experimental measurement of the production rate of \isotope{Cl}{39} in argon gas exposed at sea level \cite{rama1961cosmic}. Summing up the contributions from fast neutrons (\SI{173 \pm 71}{\atomscltn\per\kgar\per\day} through \arft\ncomb\isotope{Cl}{39}) with the alternate mechanisms listed in Table~\ref{tab:artn_total_production_rates}, one obtains a total production rate of \SI{369 \pm 76}{\atomscltn\per\kgar\per\day} which is in relatively good agreement with the experimentally measured \SI{288 \pm 29}{\atomscltn\per\kgar\per\day}. 

Thus the total cosmic ray production rate at sea-level is expected to be \SI{1048 \pm 133}{\atoms\per\kgar\per\day} for \artn, which corresponds to a cosmic ray activation rate of \SI{8.6 \pm 1.1 e-8}{\becquerel\per\kgar\per\day} and a saturated equilibrium activity of \SI{1.21 \pm 0.15 e-3}{\becquerel\per\kgar}. Note that the equilibrium activity at sea-level is lower than the measured level of \artn~in the atmosphere because the total rate in the atmosphere is dominated by the production at high altitudes, where the neutron flux is significantly higher. 

\begin{table*}[t!]
\centering
\setcellgapes{2pt} \makegapedcells \renewcommand\theadfont{\normalsize\bfseries}%
\begin{tabular}{c c r}
\hline
Reaction & Estimated \artn~Production rate  & Fraction of total \\
& [\si{\atoms\per\kgar\per\day}] &  AAr [\%]\\
\hline
\arft\ntwon\artn~+ & \multirow{2}{*}{\num{759 \pm 128}} & \multirow{2}{*}{72.3}\\
\arft\ncomb\isotope{Cl}{39} & &\\
\hline
\arft($\mu$, n)\isotope{Cl}{39} & \num{172 \pm 26} & 16.4  \\
\hline
\arft($\gamma$, n)\artn & \num{89 \pm 19} & 8.5  \\
\arft($\gamma$, p)\isotope{Cl}{39} & \num{23.8 \pm 8.7} & 2.3  \\
\hline
\arft(p, 2p)\isotope{Cl}{39} & \num{< 0.1} & $< 0.01$ \\
\arft(p, pn)\artn & \num{3.6 \pm 2.2}  & 0.3 \\
\hline
 \multirow{2}{*}{\isotope{Ar}{38}\ngamma\artn} & \num{\ll~0.1} (UAr)  & - \\
 & \num{1.1 \pm 0.3} (AAr) & 0.1\\
\hline
Total & \num{1048 \pm 133} & 100 \\
\hline
\end{tabular}
  \caption{Total cosmogenic production rates of \artn~at sea-level. The first row is the estimate from fast neutrons based on the measurement presented in this work, while the other rows are best estimates made from existing experimental data and models.}
   \label{tab:artn_total_production_rates}
\end{table*}

\subsection {Alternate \arts~Production Mechanisms}

\subsubsection*{\arft($\gamma$, 3n)\arts~and \arft(p, p3n)\arts}
We are not aware of any existing measurements of either the \arft($\gamma$, 3n)\arts~or the \arft(p, p3n)\arts~cross-section. For the production rate from gammas we used the cross-section from TALYS (shown in Figure~\ref{fig:alternate_production_cs}), while for proton induced reactions we have used the production rates obtained with the TALYS and INCL++ \arft(n, 4n)\arts~cross-sections. As above, where multiple flux or cross-section estimates exist we have used the mean and sample standard deviation of the calculated production rates as our estimates of the central value and uncertainty. We estimate a sea-level production rate of \SI{3.5 \pm 0.7}{\atoms\per\kgar\per\day} and \SI{1.3 \pm 0.4}{\atoms\per\kgar\per\day} from \arft($\gamma$, 3n)\arts~and \arft(p, p3n)\arts~respecively.

\subsubsection*{\isotope{Ar}{36} \ngamma\arts}
In AAr, \arts~can also be produced through the capture of thermal and epi-thermal neutrons on \isotope{Ar}{36}.  For the \isotope{Ar}{36}\ngamma\arts~cross-section we have used the ENDF/B-VII.I values \cite{chadwick2011endf} (shown in Figure~\ref{fig:thermal_neutrons}) which agrees well with experimental data at thermal energies \cite{mcmurtrie1950thermal, tessler2018stellar}, though recent measurements at higher energies (Maxwellian $kT\approx47$ keV) indicate that the cross-sections could be significantly lower than previous estimates \cite{tessler2018stellar}. This production channel produces a significant contribution to the total production rate of \SI{36 \pm 11}{\atoms\per\kgar\per\day}~at sea-level. Note that the abundance of \isotope{Ar}{36}~in argon from underground sources is reduced by roughly a factor of 40, and hence the production of \arts~from neutron captures in UAr is $<$\SI{1} {\atom\per\kgar\per\day}.

\subsubsection*{\isotope{Ar}{38}(n, 2n)\arts, \isotope{Ar}{38}($\gamma$, n)\arts,~and  \isotope{Ar}{38}(p, pn)\arts}
All of the direct production mechanisms of \artn~from \arft~also apply to the production of \arts~from \isotope{Ar}{38} in AAr. We are not aware of any experimental measurements of the cross-sections on \isotope{Ar}{38} except for \isotope{Ar}{38}($\gamma$, n)\arts~\cite{ehhalt1965kernphotoeffekt} which is roughly the same as the \arft($\gamma$, n)\artn~cross-section. We have therefore estimated the production rate from reactions on \isotope{Ar}{38} by taking the sum of the production rates from all direct \arft(x, y)\artn~mechanisms (\SI{679 \pm 85}{\atoms\per\kgar\per\day}) and scaling it by the abundance of \isotope{Ar}{38} in AAr (0.000629). This is perhaps an overestimation due to the fact that \isotope{Ar}{38} has a magic number of neutrons and may therefore be expected to have lower cross-sections for the ejection of a neutron than \arft, but the contribution is in any case negligible (\SI{0.43 \pm 0.05} {\atoms\per\kgar\per\day}).

\begin{table*}[t!]
\centering
\setcellgapes{2pt} \makegapedcells \renewcommand\theadfont{\normalsize\bfseries}%
\begin{tabular}{c c r r }
\hline
Reaction & Estimated \arts~Production rate  & Fraction of total \uar & Fraction of total \aar\\
& [\si{\atoms\per\kgar\per\day}] & [\%] & [\%]\\
\hline
\arft\nfourn\arts & \num{51.0 \pm 7.4} & 90.0 & 55.5\\
\hline
\arft($\gamma$, 3n)\arts & \num{3.5 \pm 0.7} & 6.1 & 3.8 \\
\hline
\arft(p, p3n)\arts & \num{1.3 \pm 0.4} & 2.3 & 1.4 \\
\hline
 \multirow{2}{*}{\isotope{Ar}{36} \ngamma\arts} & \num{0.9 \pm 0.3} (UAr)  & 1.6 & 38.9 \\
 & \num{36 \pm 11} (AAr) & & \\
\hline
 \isotope{Ar}{38}(n, 2n)\arts~+ & {$< 0.05$ (UAr)}  & \multirow{3}{*}{$< 0.1$} &  \multirow{3}{*}{0.5} \\
 \isotope{Ar}{38}($\gamma$, n)\arts~+ &  & \\
 \isotope{Ar}{38}(p, pn)\arts &  {\num{0.43 \pm 0.05} (AAr)} & &\\
\hline
 \multirow{2}{*}{Total} & \num{56.7 \pm 7.5} (UAr) &  \multirow{2}{*}{100} &   \multirow{2}{*}{100}\\
 & \num{92 \pm 13} (AAr) & \\
\hline
\end{tabular}
  \caption{Total cosmogenic production rates of \arts~at sea-level. The first row is the estimate from fast neutrons based on the measurement presented in this work, while the other rows are best estimates made from existing experimental data and models.}
   \label{tab:arts_total_production_rates}
\end{table*}

The estimates for each of these alternate production mechanisms for \arts~is summarized in Table~\ref{tab:arts_total_production_rates}. The total cosmic ray production rate at sea-level is expected to be \SI{56.7 \pm 7.5}{\atoms\per\kgar\per\day} for \arts~in underground argon and \SI{92 \pm 13}{\atoms\per\kgar\per\day} for \arts~in atmospheric argon. These numbers correspond to a cosmic ray activation rate of \SI{1.30 \pm 0.17 e-5}{\becquerel\per\kgar\per\day} for UAr and \SI{2.11 \pm 0.30 e-5}{\becquerel\per\kgar\per\day} for AAr, and a saturated equilibrium activity of \SI{6.56 \pm 0.87 e-4}{\becquerel\per\kgar}, \SI{1.06 \pm 0.15 e-3}{\becquerel\per\kgar} respectively.

%% file: discussion.tex
\section{Summary and Discussion}
We have made the first experimental measurement of the production rate of \artn~and \arts~due to fast neutron interactions in argon. This measurement was enabled by the ability to directly measure the $\beta$~and electron-capture decays that are not detectable with standard methods only sensitive to $\gamma$-rays from activation products. Including uncertainties in the cross-section models and the cosmogenic neutron flux we obtained a production rate of \SI{759 \pm 128}{\atoms\per\kgar\per\day} for \artn, and \SI{51.0 \pm 7.4}{\atoms\per\kgar\per\day} for \arts~from cosmogenic neutrons at sea-level. Combined with calculated estimates of other production mechanisms, we obtain a total cosmic ray production rate at sea-level of \SI{1048 \pm 133}{\atoms\per\kgar\per\day} for \artn, and \SI{92 \pm 13}{\atoms\per\kgar\per\day} for \arts~in atmospheric argon (\SI{56.7 \pm 7.5}{\atoms\per\kgar\per\day} for \arts~in \uar).

These results are most relevant to argon-based dark matter detectors where \artn~is the dominant background and lowering the rate of \artn~can reduce energy thresholds and improve sensitivity. The argon extracted from underground and used as the target for the DarkSide-50 experiment was measured to have an \artn~rate of \SI{7.3e-4}{\bqkg}~\cite{agnes2016results}, but it is thought that a large fraction of the residual \artn~was due to an air infiltration during the purification of the underground gas. Thus it is possible that the measured activity is only an upper limit, and the true rate of \artn~in \uar~could be as low as \SI{3e-5}{\bqkg} \cite{renshaw2018urania}. For future multi-ton dark matter detectors low levels of \artn~are even more critical, and efforts will be made to avoid any contamination of the underground argon with the atmosphere \cite{back2018device}. At this possibly lower intrinsic \artn~rate, cosmogenic activation of the underground argon (once it is extracted to the surface) is potentially a significant contributor to the overall background. The results from this paper, scaled by the cosmic ray flux at the relevant altitudes and locations, can be used to calculate the experiment-specific maximum duration \uar~can spend above ground during extraction, purification, storage, and transportation. We note that the measured sea-level activation rate of \artn~in this paper [\SI{8.6 \pm 1.1 e-8}{\becquerel\per\kgar\per\day}] is nearly an order of magnitude larger than previous estimates based on semi-empirical calculations \SIrange[fixed-exponent = -8 , range-units = brackets , scientific-notation = fixed, range-phrase = --]{0.5e-8}{1.5e-8}{\becquerel\per\kgar\per\day} \cite{aalseth2018darkside}.

Finally, these results can also be used to estimate the total equilibrium rate of \artn~and \arts~in the atmosphere. This is potentially useful for establishing baseline values for radioactive dating and nuclear activity monitoring, as well as evaluating the constancy of cosmic radiation \cite{loosli1970argon}.

%% file: acknowledgements.tex
\section{Acknowledgements}
This work was supported by the Laboratory Directed Research and Development Program at Pacific Northwest National Laboratory, a multiprogram national laboratory operated by Battelle for the U.S. Department of Energy under Contract No. DE-AC05-76RL01830, and also the U.S. Department of Energy, Office of Science, Office of Nuclear Physics under Los Alamos National Laboratory Award No. LANLE9BW. This work benefited from the use of the Los Alamos Neutron Science Center, funded by the US Department of Energy under Contract No. DE-AC52-06NA25396 and we would like to thank Steve Wender for his assistance with the beam exposure.